\documentclass[aps,twocolumn,preprintnumbers,nofootinbib,superscriptaddress]{revtex4}
\usepackage{amsmath} \usepackage{graphicx} \usepackage{amsfonts}
\usepackage{array} \usepackage{amsthm} \usepackage{bm}
\usepackage{palatino} \usepackage{mathpazo} 
\usepackage{supertabular}

\usepackage[breaklinks]{hyperref}
\usepackage{color}

\usepackage{epstopdf}

\newcommand{\be}{\begin{equation}}
\newcommand{\ee}{\end{equation}}
\newcommand{\ba}{\begin{eqnarray}}
\newcommand{\ea}{\end{eqnarray}}
\newcommand{\bal}{\begin{align}}
\newcommand{\eal}{\end{align}}

\newcommand{\bw}{\begin{widetext}}
\newcommand{\ew}{\end{widetext}}

\begin{document}

\title{\bf \Large Weak Cosmic Censorship by Overspinning the Kerr-Newman-Kasuya Black Hole with Test Particle }

\author{Ayyesha K. Ahmed}\email{aahmed.phdmath17sns@student.nust.edu.pk}
\affiliation{Department of Mathematics, School of Natural Sciences (SNS), National University
of Sciences and Technology (NUST), H-12, Islamabad, Pakistan}

\author{Azad A. Siddiqui}\email{azad@sns.nust.edu.pk}
\affiliation{Department of Mathematics, School of Natural Sciences (SNS), National University
of Sciences and Technology (NUST), H-12, Islamabad, Pakistan}

\begin{abstract}
We check the weak cosmic censorship conjecture by attempting to overspin the Kerr-Newman-Kasuya black hole with test particle. The study suggests that the Kerr-Newman-Kasuya black hole can be over-spun showing the violation of cosmic censorship conjecture.
\end{abstract}

\pacs{04.40.Dg, 95.30.Sf, 04.50.Gh}



\maketitle

\section{Introduction}
The gravitational collapse results in the formation of spacetime singularities. These singularities are hidden behind the event horizons of black holes and can not be seen by the distant observers. This was the conjecture proposed by Roger Penrose in $1969$ \cite{A}, called the weak cosmic censorship conjecture (WCCC). There are certain methods to check the validity of this conjecture. Wald has tested its validity through Gedanken experiment in $1974$ \cite{B}. He threw a test particle into the black hole and observe whether the event horizon of the black hole was destroyed or not. Since the accretion of particle increases energy and angular momentum of the black hole so there is a possibility of the formation of naked singularity. This leads to the concept of over-charging or over-spinning of black holes. Hubney found that it is possible to overcharge the near-extremal charged black hole by capturing the charged test particles \cite{C}. This idea was further extended by Jacobson and Sotiriou and they found that over-spinning is possible for the Kerr black hole \cite{D}. There are many further studies, in which it is shown that over-spinning is possible \cite{E,F,G,H,I,J}.

\par There are several studies which relate the over-spinning/over-charging and cosmic censorship conjecture. Wald discussed that if a particle is to destroy horizon of a black hole then it must not be captured by the black hole due to the repulsive force \cite{B}. Later on, many studies showed that overspinning a black hole is possible such as Kerr-Newman black hole \cite{B}, BTZ black holes \cite{G}, Higher dimensional Myers-Perry rotating black hole \cite{J}. Further Gwak explored that both extremal and non-extremal Kerr anti de-Sitter black holes cannot be over spun \cite{K}. There is a series of work which shows that five dimensional rotating black holes \cite{L}, BTZ black hole \cite{M,N}, higher dimensional charged adS black holes \cite{O,P} cannot be over-spun and over-charge by test particles. This suggests that the weak cosmic censorship conjecture is preserved for these black holes. In these studies, back-reaction and self-force effects are ignored. However, these effects are considered in references \cite{Q,R,S,T,U,V}.

\par Generally, it is believed that charged black holes do not exist but due to the gravitational collapse magnetized black holes are formed having equal and opposite charges on them \cite{N1}. This gives rise to the electric and magnetic monopoles. Dyon is a pole that possess both electric and magnetic charges. The Kerr-Newman-Kasuya (KNK) is an interesting case of a rotating dyon \cite{1a}. It consists of four physical parameters i.e. mass $M$, rotation parameter $a$, electric charge $Q_{e}$ and magnetic charge $Q_{m}$. This dyon prevails in the Abelian theory \cite{1a}.

\par The main objective here is to check the WCCC by overspinning the KNK black hole. The paper is organized as follows: In Section II a brief discussion of the KNK black hole is given. In Section III thermodynamics of the KNK black hole is discussed. In the next section the details of overspining the black hole are given, and in the end a brief discussion and conclusion is given. Throughout the study, the metric signatures is $(-,+,+,+)$ and $G=c=1$.

\section{The KNK Black Hole}	
\noindent The Lagrange density which describes the KNK black hole is given by \cite{N11}
\begin{eqnarray}\label{N1}
L&=&\sqrt{-g}\Big[-\Big(\frac{1}{16\pi}\Big)R-\frac{1}{4}g^{\mu\rho}g^{\nu\sigma}F_{~\mu\nu}F_{~\rho\sigma}\Big],
\end{eqnarray}
with
\begin{eqnarray}\label{N2}
F_{~\mu\nu}&=&\partial_{\mu}A_{\nu}-\partial_{\nu}A_{\mu}+^{\ast}G_{\mu\nu},
\end{eqnarray}
where $^{\ast}G_{\mu\nu}$ is the Dirac String term \cite{N22}. The field equations are given by
\begin{eqnarray}
R^{\mu\nu}-\frac{1}{2}g^{\mu\nu}R&=&8\pi T^{\mu\nu},\label{N3}\\
\partial_{\nu}\Big(\sqrt{-g}g^{\mu\rho}g^{\nu\sigma}F_{~\rho\sigma}\Big)&=&0,
\label{N4}
\end{eqnarray}
where the energy momentum tensor, $T^{\mu}_{~\nu}$, is given by
\begin{eqnarray}
T^{\mu}_{~\nu}&=&-\Big(g^{\mu\alpha}g^{\rho\beta}F_{~\alpha\beta}F_{~\nu\rho}-\frac{1}{4}\delta^{\mu}_{~\nu}g^{\rho\alpha}g^{\sigma\beta}F_{~\rho\sigma}F_{~\alpha\beta}\Big).
\end{eqnarray}
For the detailed discussion of action and field equations for KNK black hole, one may refer to \cite{N11}. The line element of KNK black hole in Boyer-Lindquist coodinates is given by \cite{N11,1a}
\begin{eqnarray}\label{1}
ds^{2}&=&-\frac{\Delta}{\Sigma}\Big(dt-a\sin^{2}\theta d\phi\Big)^{2}+\frac{\Sigma}{\Delta}dr^{2}+\Sigma d\theta^{2}\nonumber\\
&+&\frac{\sin^{2}\theta}{\Sigma}\Big[(r^{2}+a^{2})d\phi-a dt\Big]^{2},
\end{eqnarray}
where
\begin{eqnarray}
\Delta&=&r^{2}-2Mr+a^{2}+Q_{e}^{2}+Q_{m}^{2},\label{2}\\
\Sigma&=&r^{2}+a^{2}\cos^{2}\theta.\label{3}
\end{eqnarray}
Here $M$ is the mass of the black hole, $a$ is the rotation parameter defined as $a=J/M$ and $Q_{e}$ and $Q_{m}$ denote the electric and magnetic charges of the black hole respectively. The vector potential of the KNK black hole \cite{N11} is expressed as
\begin{eqnarray}
A_{t}&=&\frac{1}{\Sigma}\Big[-Q_{e}r+Q_{m}a\cos\theta\Big],\label{15}\\
A_{r}&=&A_{\theta}=0,\label{16}\\
A_{\phi}&=&\frac{Q_{e}}{\Sigma}ra\sin^{2}\theta+Q_{m}\Big[1-\frac{r^{2}+a^{2}}{\Sigma}\cos\theta\Big].\label{17}
\end{eqnarray}
If the parameters $a=Q_{e}=Q_{m}=0$, then the KNK black hole in Eq. (\ref{1}) reduces to the Schwarzschild black hole and for $a=Q_{m}=0$, we get the Reissner-Nordstrom black hole. Likewise, for $Q_{e}=Q_{m}=0$, it reduces to the Kerr black hole and for $Q_{m}=0$, it reduces to the Kerr-Newman black hole. The inner and outer horizons are obtained by putting $\Delta=0$ and are given as
\begin{eqnarray}\label{4}
r_{\pm}&=&M\pm\sqrt{M^{2}-(a^{2}+Q_{e}^{2}+Q_{m}^{2})}.
\end{eqnarray}
The outer horizon $r_{+}$ corresponds to the event horizon. For the extremal black hole both inner and outer horizons concide which gives
\begin{eqnarray}\label{5}
M^{2}&=&a^{2}+Q_{e}^{2}+Q_{m}^{2},
\end{eqnarray}
and hence the degenerate horizon \cite{self} for extremal case is located at $r_{ext}=M$. The naked singularity occurs when $M^{2}<a^{2}+Q_{e}^{2}+Q_{m}^{2}$.

\section{Thermodynamics of the KNK Black Hole}
\noindent Consider a particle of mass $m$ and charge $q$ falling into the KNK black hole. When the particle falls into the black hole, it must obey the first law of black hole thermodynamics. Area of the KNK black hole is
\begin{eqnarray}\label{6}
A_{+}&=&4\pi(r_{+}^{2}+a^{2}),
\end{eqnarray}
while its surface gravity is given as
\begin{eqnarray}
\kappa_{+}&=&\frac{1}{2(r_{+}^{2}+a^{2})}\frac{d\Delta}{dr}\Big|_{r=r_{+}},\nonumber\\
&=&\frac{1}{2}\frac{(r_{+}-r_{-})}{r_{+}^{2}+a^{2}}.\label{7}
\end{eqnarray}
The Hawking temperature reads
\begin{eqnarray}\label{8}
T_{+}&=&\frac{\kappa_{+}}{2\pi}=\frac{r_{+}-r_{-}}{4\pi(r_{+}^{2}+a^{2})},
\end{eqnarray}
and its entropy is calculated as
\begin{eqnarray}\label{9}
S_{+}&=&\frac{A_{+}\kappa}{4}=\frac{(r_{+}^{2}+a^{2})}{8},
\end{eqnarray}
where $\kappa=1/8\pi$. The angular velocity of the KNK black hole is
\begin{eqnarray}\label{10}
\Omega_{+}&=&-\frac{g_{t\phi}}{g_{\phi\phi}}\Big|_{r=r_{+}}=\frac{a}{r_{+}^{2}+a^{2}},
\end{eqnarray}
and the electric potential is \cite{2a}
\begin{eqnarray}\label{11}
\Phi_{+}&=&\frac{Q_{e}r_{+}-Q_{m}a}{r_{+}^{2}+a^{2}}.
\end{eqnarray}

\noindent Using the expressions of thermodynamic quantities given by Eq. (\ref{6})-(\ref{11}), the first law of thermodynamics for the KNK-BH is given in reference \cite{2a} as
\begin{eqnarray}\label{12}
dM&=&T_{+}dS_{+}+\Omega_{+}dJ+\Phi_{+}dQ_{e}.
\end{eqnarray}
\indent

\section{Weak Cosmic Censorship with Test Particle}
\noindent In this section, we check the weak cosmic censorship conjecture by throwing a massive charged rotating test particle into the black hole. We check whether the event horizon of the KNK black hole can be destroyed or not. Motion of the test particle \cite{3a} is given by the equation
\begin{eqnarray}\label{13}
\ddot{x}^{\mu}+\Gamma^{\mu}_{~\alpha\beta}\dot{x}^{\alpha}\dot{x}^{\beta}&=&\frac{q}{m}F^{\mu\nu}\dot{x}_{\nu},
\end{eqnarray}
and the corresponding Lagrangian is
\begin{eqnarray}\label{14}
\mathcal{L}&=&\frac{m}{2}g_{\alpha\beta}\dot{x}^{\alpha}\dot{x}^{\beta}+q A_{\mu}\dot{x}^{\mu}.
\end{eqnarray}
The energy and angular momentum of the particle are
\begin{eqnarray}
\delta E&=&-P_{t}=-\frac{\partial \mathcal{L}}{\partial\dot{t}}=-m\Big(g_{tt}\dot{t}+g_{t\phi}\dot{\phi}\Big)-qA_{t},\label{18}\\
\delta J&=&P_{\phi}=\frac{\partial \mathcal{L}}{\partial\dot{\phi}}=m\Big(g_{t\phi}\dot{t}+g_{\phi\phi}\dot{\phi}\Big)+qA_{\phi}.\label{19}
\end{eqnarray}
Now we find $\delta E$ and $\delta J$ and check whether or not the particle violates the cosmic censorship conjecture. Since the four velocity, $u^{\mu}$, of the particle is time-like, therefore, the normalization condition, $u^{\mu}u_{\mu}=-1$, is
\begin{eqnarray}\label{20}
g^{\mu\nu}(P_{\mu}-qA_{\mu})(P_{\nu}-qA_{\nu})&=&-m^{2}.
\end{eqnarray}
Using Eqs. (\ref{18}) and (\ref{19}) in Eq. (\ref{20}) we get
\begin{widetext}
\begin{eqnarray}\label{21}
\delta E&=&\frac{g_{t\phi}}{g_{\phi\phi}}\Big(qA_{\phi}-\delta J\Big)-qA_{t}+\sqrt{\Big(\frac{g_{t\phi}^{2}-g_{tt}g_{\phi\phi}}{g_{\phi\phi}^{2}}\Big)\Big[(\delta J-qA_{\phi})^{2}+m^{2}g_{\phi\phi}\Big(1+g_{rr}\dot{r}^{2}+g_{\theta\theta}\dot{\theta}^{2}\Big)\Big]}.
\end{eqnarray}
\end{widetext}
For motion of the particle to be future directed, $\frac{dt}{d\tau}>0$ which is equivalent to the following requirement
\begin{eqnarray}\label{22}
\delta E&>&\Big[\frac{g_{t\phi}}{g_{\phi\phi}}\Big(qA_{\phi}-\delta J\Big)-qA_{t}\Big]\Big|_{r=r_{+}},\nonumber\\
&=&-\frac{g_{t\phi}}{g_{\phi\phi}}\delta J\Big|_{r=r_{+}}+\frac{g_{t\phi}}{g_{\phi\phi}}qA_{\phi}\Big|_{r=r_{+}}-qA_{t}\Big|_{r=r_{+}},\nonumber\\
&=&\Omega_{+}\delta J-q\Big[A_{t}-\frac{g_{t\phi}}{g_{\phi\phi}}A_{\phi}\Big]\Big|_{r=r_{+}}.
\end{eqnarray}
At the equatorial plane, the above equation reduces to
\begin{eqnarray}\label{23}
\delta E&>&\Omega_{+}\delta J+q\Big[\frac{aQ_{m}-Q_{e}r_{+}}{Q_{e}^{2}+Q_{m}^{2}-2Mr_{+}}\Big],
\end{eqnarray}
which implies
\begin{eqnarray}\label{25}
\delta J&<&\frac{\delta E}{\Omega_{+}}-\frac{q}{\Omega_{+}}\Big(\frac{aQ_{m}-Q_{e}r_{+}}{Q_{e}^{2}+Q_{m}^{2}-2Mr_{+}}\Big).
\end{eqnarray}
Using Eqs. (\ref{2}) and (\ref{10}) at the horizon, $r_{+}$, the above inequality (\ref{25}) can be expressed as
\begin{eqnarray}\label{25a}
\delta J&<&\frac{(r_{+}^{2}+a^{2})\delta E}{a}+\frac{q(aQ_{m}-Q_{e}r_{+})}{a}=\delta J_{max}.
\end{eqnarray}
Using the Smarr Mass Formula \cite{5a}, mass of the KNK black hole is calculated as
\begin{eqnarray}\label{31}
M^{2}&=&2S+\frac{J^{2}}{8S}+\frac{Q_{e}^{2}+Q_{m}^{2}}{2}+\frac{(Q_{e}^{2}+Q_{m}^{2})^{2}}{32S}.
\end{eqnarray}
From Eqs. (\ref{12}) and (\ref{31}) we have
\begin{eqnarray}
Q_{e}&=&\frac{1}{3}+\frac{2\sqrt{1-3Q_{m}^{2}}}{3}\cos\Big[\frac{1}{3}\cos^{-1}\frac{\Upsilon}{\Theta}\Big],\label{33}
\end{eqnarray}
where,
\begin{eqnarray}
\Upsilon&=&18Q_{m}^{2}-54MQ_{m}a+2,\label{34}\\
\Theta&=&2(1-3Q_{m}^{2})^{3/2}.\label{35}
\end{eqnarray}
The minimum of the metric function is given as
\begin{eqnarray}\label{26}
\Delta_{min}&=&a^{2}-M^{2}+Q_{e}^{2}+Q_{m}^{2}.
\end{eqnarray}
Now differentiating Eq. (\ref{26}) w.r.t. mass, $M$, and angular momentum, $J$, using Eq. (\ref{33}) we have
\begin{eqnarray}
\Big(\frac{\partial\Delta_{min}}{\partial M}\Big)&=&-2M,\label{36}\\
\Big(\frac{\partial\Delta_{min}}{\partial J}\Big)&=&\frac{2a}{M^{3}}.\label{37}
\end{eqnarray}
When the test particle with energy $\delta E$ and momentum $\delta J$ is absorbed, the black hole parameters become
\begin{eqnarray}
M\rightarrow M^{\prime}&=&M+dM,\label{27}\\
J\rightarrow J^{\prime}&=&J+dJ,\label{28}\\
Q_{e}\rightarrow Q_{e}^{\prime}&=&Q_{e},\label{29}\\
Q_{m}\rightarrow Q_{m}^{\prime}&=&Q_{m}.\label{30}
\end{eqnarray}
Therefore, $\Delta_{min}\rightarrow\Delta^{\prime}_{min}$ i.e.
\begin{eqnarray}
\Delta^{\prime}_{min}&=&\Delta^{\prime}_{min}(M+dM,J+dJ,Qe,Qm),\nonumber\\
&=&\Delta_{min}+\Big(\frac{\partial\Delta_{min}}{\partial M}\Big)dM+\Big(\frac{\partial\Delta_{min}}{\partial J}\Big)dJ.\label{N11}
\end{eqnarray}
Using the expressions (\ref{36}) and (\ref{37}) in Eq. (\ref{N11}) we have
\begin{eqnarray}
\Delta^{\prime}_{min}&=&-(M^{2}-a^{2}-Q_{e}^{2}-Q_{m}^{2})-2MdM+\frac{2a}{M^3}dJ.~~~~~~~~\label{38}
\end{eqnarray}
Now, to over-spin the black hole, after the absorption of the particle, the minimum of the metric function $\Delta^{\prime}_{min}$ should be positive i.e. $\Delta^{\prime}_{min}>0$. To the first order this becomes
\begin{eqnarray}\label{N22}
-(M^{2}-a^{2}-Q_{e}^{2}-Q_{m}^{2})-(2M-4r_{+})\delta M+\frac{2a}{M}\delta J>0,~~~~~~
\end{eqnarray}
which implies
\begin{eqnarray}\label{39}
\delta J&>&\frac{M(M^{2}-a^{2}-Q_{e}^{2}-Q_{m}^{2})}{2a}\nonumber\\
&+&\frac{(M^{2}-2Mr_{+})\delta E}{a}=\delta J_{min}.~~~
\end{eqnarray}
Using Eq. (\ref{4}), Eqs. (\ref{25a}) and (\ref{39}), after simplifying together imply
\begin{widetext}
\begin{eqnarray}\label{40}
\frac{MX^{2}}{2a}-\frac{M(M+X)\delta E}{a}<\delta J<\frac{\Big[(M+X)^{2}+a^{2}+M(M+X)\Big]\delta E}{a}-\frac{q}{a}\Big[Q_{e}(M+X)-aQ_{m}\Big],
\end{eqnarray}
\end{widetext}
where,
\begin{eqnarray}
X&=&\sqrt{M^{2}-(a^{2}+Q_{e}^{2}+Q_{m}^{2})}.
\end{eqnarray}
The overspinning is possible when the inequality (\ref{40}) is satisfied showing the violation of the cosmic censorship conjecture. (For example, for the the parameter values $M=2$, $\delta E=a=Q_{m}=Q_{e}=1$ and $q=0.1$, one can see that the inequality holds.)


\section{Discussion and Conclusions}
In this paper we have studied the weak cosmic censorship conjecture for the Kerr-Newman-Kasuya black hole. For this purpose, we have tried to over-spin the black hole with the help of test particle. Our study shows that the Kerr-Newman-Kasuya black hole can be over-spun by the test particle, showing that the weak cosmic censorship conjecture is violated. It will be interesting to extend our study, for instance, by taking into account the back-reaction effects or by overcharging the Kerr-Newman-Kasuya black hole.

\end{document}